\documentclass[preprint,showpacs,preprintnumbers,amsmath,amssymb]{revtex4}
\usepackage{graphicx}    
\usepackage{dcolumn}     
\usepackage{bm}          
\usepackage{bbold}
\usepackage{phonetic}
\newcommand{\be}{\begin{equation}}
\newcommand{\ee}{\end{equation}}
\newcommand{\bes}{\begin{equation*}}
\newcommand{\ees}{\end{equation*}}
\newcommand{\bea}{\begin{eqnarray}}
\newcommand{\eea}{\end{eqnarray}}
\newcommand{\bean}{\begin{eqnarray*}}
\newcommand{\eean}{\end{eqnarray*}}
\newcommand{\ba}{\begin{array}}
\newcommand{\ea}{\end{array}}

\newcommand{\Per}{\mathrm{Per}}

\newcommand{\dagga}{{\phantom{\dagger}}}
\newcommand{\eqn}[1]{(\ref{#1})}

\begin{document}
\title{\Large{How localized bosons manage to become superfluid}}
\author{Luca Dell'Anna$^{1}$ and Michele Fabrizio$^{2,3}$}
\affiliation{$^1$ Dipartimento di Fisica ``G. Galilei'', Universit\`a di Padova, 35131, Italy\\
$^2$ International School for Advanced Studies, SISSA, 34136 Trieste, Italy\\
$^3$ International Centre for Theoretical Physics, ICTP, 34151 Trieste, Italy}
\date{\today}
\begin{abstract}
We show that the many-body wavefunction built as a permanent of 
localized non-orthogonal single-particle states 
can describe bosonic superfluidity. 
The criterium for the transition is expressed in terms of 
the properties of the matrix of the overlaps between the single-particle 
wavefunctions. 
We apply this result to study the superfluid-to-Bose glass transition 
in a disordered Bose-Hubbard model through a very simple variational 
wavefunction. 
We finally consider a further quantity, the bipartite entanglement entropy, 
which also provides a good estimator for the  
superfluid-to-Bose glass transition. 
\end{abstract}
\maketitle
\section{Introduction}
The study of the interplay between disorder and interaction has recently 
received novel impulse 
by the experimental realizations of correlated and disordered systems 
through optical 
lattices.\cite{cirac, bloch} Particular interest has been devoted to trapped 
bosonic atoms  
in the presence of on-site disorder, which can be simulated experimentally 
using speckles or 
superimposing incommensurate laser beams.\cite{billy, roati, fallani}
The great opportunity offered by this kind of experiments is that 
not only disorder can be easily tuned
but also the interaction among the atoms, this latter  
by means of the Feshbach's resonance technique.\cite{chin}
These experiments have motivated a renewed theoretical activity on the phase 
diagram of the disordered 
Bose-Hubbard model
\cite{schollowck, lee, damski, proko, buonsante2, troyer, gurarie, hofstetter,carrasquilla}, an 
old\cite{giamarchi-1, giamarchi, fisher, fm, krauth, pai} but still 
debated issue.
This model is supposed to exhibit three different phases. 
Besides the Mott-insulating phase, 
occurring when interaction is strong and the particle density is commensurate,  
and the superfluid phase, stable for weak interaction, in the presence 
of disorder an additional 
phase has been predicted to occur,\cite{fisher} the so-called Bose glass. 
Such a disorder driven phase 
is supposed to be insulating but, unlike the Mott insulator, compressible.
In the absence of interaction, the Bose glass is simply the state obtained by 
condensing all
bosons in the lowest energy eigenstate of the single-particle 
disordered Schr{\"o}dinger equation, 
which is supposed to lie in the Lifshits's tail of localized wavefunctions. 
This picture is however unstable to whatever weak interaction, since a 
macroscopy occupancy of a finite region in real space would be 
energetically unbearable. 
In fact, a more realistic view of a Bose glass is that of 
disconnected superfluid droplets, 
where coherent inter-droplet tunneling is inhibited by the 
Anderson localization phenomenon, 
hence the insulating behavior, yet transferring electrons between 
different droplets is costless, hence 
the finite compressibility.      

In this paper, we investigate the superfluid-to-Bose glass transition 
by means of a very simple 
variational wavefunction. It consists of a permanent of non-orthogonal 
single-particle wavefunctions that 
are determined in a variational manner. Although the approach is not as rigorous 
as e.g. quantum Monte 
Carlo,\cite{lee,proko,troyer,carrasquilla} nevertheless it provides a very 
transparent description of 
the transition and of the same Bose glass. Indeed, all coherence properties 
of the wavefunction are hidden 
in the single-particle wavefunctions that build the permanent. 
In particular, since these wavefunctions need not to be orthogonal, 
it is their overlap matrix that seems to play an important 
role. Actually one can derive an approximate criterium for the permanent to have 
macroscopic condensation at zero momentum, hence superfluidity, 
which involves just the overlap matrix. 
We find that this criterium agrees well with the more rigorous one 
based on the superfluid stiffness.     

Finally, in the last part of the paper we study the single 
site von Neumann entropy, which could also be used to identify the 
superfluid-to-Bose glass transition. The quantum entanglement 
applied to many body systems has attracted lot of theoretical interest in 
recent years (see for instance Refs. \cite{buonsante, amico, haque, katsura} 
and references therein). So far, however, the entanglement witnesses have 
never been used in the context of disordered Hubbard model. 
We show that, within our variational approach, the Bose glass seems to be 
characterized by a finite probability of the 
single-site von Neumann entropy to have zero value. In other words, as disorder increases, 
the entropy distribution gets broaden and broaden until the probability to have 
zero entropy becomes finite, which we identify as the onset of Bose glassiness. 
This criterium is qualitatively good as far as we deal with small system size 
since the zero entropy is the signature of vanishing charge fluctuations in some regions 
of the sample which becomes then easily disconnected. 
A more quantitative analysis for large systems would involve the study of the entropy 
distribution and the appearance of the bimodal profile in agreement with the droplet scenario. 
Our conjecture is that the weights and the positions of the two modes at the transition should be related to the percolation threshold.
 
The paper is organized as follows. In Sec.~\ref{sec.path} we present a criterium for the onset of superfluidity that is based   
on the matrix of the overlaps between the single particle wavefunctions that built the variational many-body wavefunction.
In Sec.~\ref{sec.model} we introduce the 
disordered Bose-Hubbard model and in Sec.~\ref{sec.method} we describe the variational 
many-body wavefunction that we study. Sec.~\ref{sec.entropy} is devoted to studying the bipartite 
entanglement entropy as another tool to identify the superfluid-to-Bose glass transition. 
Final conclusions are summarized in Sec.~\ref{sec.conclusions}.

\section{Overlap matrix criterium}
\label{sec.path}
Let us imagine a Hartree-Fock-like variatonal approach 
to the problem of disordered and interacting bosons that amounts to search 
for the permament that minimizes the total energy. Unlike conventional 
Hartree-Fock theory for fermions, such an approximation for bosons does not 
lead to significant simplifications with respect to exact numerical simulations 
because permanents are extremely difficult to handle with. 
Nonetheless, let us assume we have successfully performed the calculation 
and found the optimal permanent for $N$ particles, which can be written as      
\be
\label{wavef}
|\Psi\rangle=
\prod_{\alpha=1}^N\, \left(\sum_i\psi_{\alpha i}\,b_i^\dagger\right)\,|0\rangle,
\ee
in terms of a set of single particle wavefunctions $\psi_\alpha$, with amplitude 
$\psi_{\alpha i}$ at site $i$. In \eqn{wavef} $b^\dagger_i$ creates a boson at 
site $i$. 
Unlike Slater determinants, the wavefunctions that built a permanent need 
not to be orthogonal one to each other, so we expect   
the overlap matrix $\Omega$ to have non-zero 
off-diagonal elements $\Omega_{\alpha\beta} = \sum_i\,\psi^\dagga_{\alpha i}
\psi_{\beta i}^*$. For instance, if all bosons condense into a single state, then 
all $\psi_\alpha$ are equal and 
$\Omega_{\alpha\beta} = 1,~\forall \alpha,\beta\in [1,N]$. In the generic case where 
$\psi_\alpha$ are distinct, we may wonder whether wavefunction (\ref{wavef}) 
describes a superfluid. In what follows we derive a simple criterium that is 
based on the properties of the overlap matrix.       

One can verify that the norm of \eqn{wavef} can be written as an 
integral over classical variables as\cite{fisher}  
\be
\label{norm}
\langle \Psi|\Psi\rangle=\int \prod_\alpha \frac{d \xi_\alpha d \xi_\alpha^*}{\pi}e^{-{\cal{S}}(\xi,\xi^\dagger)}
\ee
with the following action (see Appendix~\ref{action})
\be
\label{act}
{\cal{S}}(\xi,\xi^\dagger)=\xi^\dagger(\Omega-{\mathbb{1}})^{-1}\xi-\sum_\alpha
\,\ln\Big(1+|\xi_\alpha|^2\Big).
\ee
We assume that the main contribution to the integral comes from 
the saddle point, i.e. the solution of 
\be
\xi_\alpha = \sum_{\beta}\, \left(\Omega_{\alpha\beta}-\delta_{\alpha\beta}\right)
\,
\frac{\displaystyle 1}
{\displaystyle 1+|\xi_\beta|^2}
\,\xi_\beta. \label{sp-eq}
\ee
The above equation implies that finite values of $\xi_\alpha$ appear in groups, 
or equivalently that $\Omega$ is a block matrix.  
In the presence of disorder this is suggestive of the existence 
of clusters occupied by bosons whose wavefunctions mutually overlap. 
Because of interaction, a cluster can not accomodate all particles unless it 
covers all the system, which would correspond also to superfluidity. If we 
linearize \eqn{sp-eq}, we find that the condition for the apperance of a 
cluster reads    
\be
\xi_\alpha 
=\sum_{\beta}\, \left(\Omega_{\alpha\beta}-\delta_{\alpha\beta}\right)\,\xi_\beta,
\label{sp}
\ee
which corresponds to a block of $\Omega$ that acquires an eigenvalue greater 
than two. As we mentioned, this is still not the condition for superfluidity. 
The latter rather implies that a block in $\Omega$  
should grow, or several blocks should merge, i.e. start to overlap, 
till a percolating cluster emerges. This condition is likely to be equivalent 
to an eigevalue of $\Omega$ that grows with the number of bosons $N$.    
We finally mention that the saddle point approximation that we used is 
rigrously valid only if blocks are big enough.

\subsection{Density matrix and overlap condition}
A better and more transparent criterium to detect a long range order is 
to resort to the definition of the density matrix,
\be
C_{ij} =\langle \Psi|b_i^\dagger b^\dagga_j|\Psi\rangle.
\ee
Off-diagonal long-range order implies that $C_{ij}$ is finite for 
$|i-j|\to \infty$. 
Within the path integral formulation, 
using Eqs.~(\ref{ba}),~(\ref{hausdorff}), one can verify that $C_{ij}$ can 
be written as
\be
C_{ij} =\sum_{\alpha\beta}\psi_{i\alpha}^\dagger\,
\langle \xi_\alpha \xi_\beta^\dagger\rangle\,\psi_{\beta j}
\ee
where $\langle ... \rangle$ is now the average weighted by 
$\mathrm{e}^{-{\cal S}}$, with the action $\cal{S}$ given by Eq.~(\ref{act}). 
If the saddle point of the action is characterized by finite $\xi_\alpha$, one 
could be tempted to set 
\be
\langle \xi_\alpha \xi_\beta^\dagger\rangle 
\simeq \langle \xi_\alpha\rangle\langle \xi_\beta^\dagger\rangle.
\label{assume}
\ee
This is not fully correct. Indeed, if $\Omega$ is a block matrix, 
within each block 
only the relative phases of the $\xi_\alpha$ are fixed, while an overall phase 
is still free and has to be integrated out. This implies that \eqn{assume} is 
correct only if $\alpha$ and $\beta$ are within the same block, otherwise the 
relative phase between the two blocks will suppress the average. This suggests 
that off-diagonal long-range order sets up only if a single block percolates. 
More rigorously, we define 
\be
\xi_i=\sum_{\alpha}\psi_{i\alpha}^\dagger \xi_\alpha,
\ee
through which the density matrix reads 
\be
C_{ij}=\langle \xi_i \,\xi_j^\dagger\rangle.
\ee
The saddle point equation in terms of $\xi_i$ is 
\be
2\,\xi_{i}=\sum_j {\sf{O}}_{ij}\, \xi_{j},\label{xi_i}
\ee
where 
\be
{\sf{O}}_{ij} =\sum_{\alpha}\,\psi_{\alpha i}^* \,\psi_{\alpha j}^\dagga,\label{O}
\ee
is the density matrix of the wavefunctions. An extreme superfluid solution 
identified by $\xi_i=\xi,~\forall i$ could be stabilized if 
\be
\label{cal_F}
{\cal F}\equiv \frac{1}{N_s}\sum_{ij}{\sf{O}}_{ij} \ge 2,
\ee
where $N_s$ is the number of sites. 
\subsection{An example: the bosonic crystal}
As a simple application of the previous results we shall now investigate 
the possibility that a permanent that describes at the 
``Hartree-Fock'' level a Bose-Wigner crystal, could be also   
superfluid, actually a supersolid.  
Let us consider a commensurate density $N/N_s<1$ of interacting bosons 
on a lattice with $N_s$ sites in the absence of disorder. If the repulsion 
is sufficiently strong and long ranged, we may imagine that the best 
variational permament wavefunction describes a bosonic 
superlattice with Bravais vectors 
\be
{\sf R}=a(m_1,m_2,...,m_d),
\ee 
where $a$ is the superlattice parameter. We  
write the Wannier single-particle wavefuncions 
that correspond to the Bose-Wigner crystal as
\be
\psi_{\sf{R}}(\mathbf{R_i})=\sqrt{\frac{1}{N_s}}\sum_{\bf k} u_{\bf k} 
e^{i{\bf k}(\sf{R}-\mathbf{R_i})}, 
\ee
where $\mathbf{k}$ runs within the Brillouin zone of the original lattice, 
$\mathbf{R}_i$ spans all lattice sites while $\sf{R}$ only the 
superlattice ones. The permanent is therefore 
\be
|\Psi\rangle = \prod_{\sf{R}}\, 
\Big(\sum_i\,\psi_{\sf{R}}(\mathbf{R_i})\,b^\dagger_i\Big)\,|0\rangle.
\ee
In this case the overlap matrix is 
\be
\Omega_{\sf{R}\sf{R}'}=
\frac{1}{N_s}\sum_{\bf k} |u_{\bf k}|^2 e^{i{\bf k}(\sf{R}-\sf{R}')}. 
\ee
We assume for simplicity that the wavefunction is gaussian, 
\be
u_{\bf k}=\frac{1}{v}\left(\frac{\ell}{\sqrt{\pi}}\right)^d e^{-\ell^2|{\bf k}|^2/4\pi^2},
\ee 
where $v$ is the volume of unit cell and $\ell$ the localization length, so  
that the overlap matrix has the simple expression 
\be
\Omega_{\sf{R}\sf{R}'}=e^{-\pi^2\frac{|\sf{R}-\sf{R}'|^2}{\ell^2}},
\ee
and depends only on the distance, $\Omega_{\sf{R}\sf{R}'} = \Omega_{\sf{R}-\sf{R}'}$. 
From the saddle point equation we find that the condition 
to have superfluidity is simply
\be
\sum_{\sf{R}} \Omega_{\sf{R}} = \sum_{\sf R} e^{-\pi^2\frac{|{\sf R}|^2}{\ell^2}} 
\ge 2.\label{cond}
\ee
By means of the Jacobi 
theta function $\theta_3(0|x)=\sum_{m=-\infty}^{\infty} x^{m^2}$, Eq.~\eqn{cond} 
means 
\be
\theta_3\left(0| e^{-\pi^2 a^2/\ell^2}\right)\ge \sqrt[d]{2}.
\label{theta}
\ee
The condition Eq.~(\ref{theta}) fixes the critical overlap between the 
Wannier functions, parametrized by the ratio $\ell/a$, above 
which bosons condense at zero 
momentum, i.e. the many-body wavefunction describes a supersolid.   

In the next section we shall consider the Bose-Hubbard model 
in the presence of disorder which causes now the localization 
of the single-particle wavefunctions. 
We are going to see that, also in that case, in spite of the 
Anderson localized nature of the single-particle states 
obtained variationally, the many-body 
wavefunction can be superfluid.

\section{The model}
\label{sec.model}
We consider a system of interacting bosons on a disordered $2$-dimensional 
lattice with $N_s=L^2$ sites, described by the following Bose-Hubbard hamiltonian
\be
H=-\frac{t}{2} \sum_{\langle ij\rangle}\, \bigg(b_i^\dagger b^\dagga_j + H.c.\bigg) 
+ \sum_i\, \epsilon_i n_i +\frac{U}{2} \sum_i\, n_i(n_i-1),\label{Ham}
\ee
where $b_i^\dagger$ ($b^\dagga_i$) creates (annihilates) a boson at site $\mathbf{R}_i$, 
$\langle ij\rangle$ denotes the sum over all pairs of neighboring sites, 
$n_i=b_i^\dagger b_i^\dagga$ is the boson local density, and, finally, 
$\epsilon_i$ are random on-site energies uniformly distributed 
between $-\Delta$ and $\Delta$.

Since the seminal works of Giamarchi and Schulz\cite{giamarchi-1, giamarchi} and 
Fisher {\sl et al.}\cite{fisher}, the Hamiltonian \eqn{Ham} has been 
studied with a variety of techniques, mainly in one and two dimensions. 
More recently, highly sophisticated numerical 
simulations\cite{krauth, pai,schollowck, lee, proko, troyer, gurarie,carrasquilla} have been performed to 
uncover the full phase diagram and settle up some debated issues,
like the possibility of a direct superfluid-Mott insulator transition. 
The results we are going to present are by no means 
comparable in accuracy with the aforementioned numerical simulations. Our 
main purpose is not to compete with those simulations, but just to provide 
an interpretation of the phase diagram in terms of a simple 
Hartree-Fock-like single-particle picture. 
 
\section{The method}
\label{sec.method}
The simplest way to deal with interacting electrons is the Hartree-Fock approximation, which amounts to 
search for the best wavefunction within the subspace of Slater determinants.  
This approximation reduces the complicated many-body problem 
to a single-particle one with an effective potential generated 
by all other particles. Even more realistic approaches, 
like the Density-Functional theory in the 
Local Density approximation, eventually reduces to 
the self-consistent solution of a 
single-particle Schr{\"o}dinger equation. 
The great advantage is that a single-particle description 
is very intuitive and, although it could be too na\"ive in many cases, 
at least it is a simple 
starting point for more complicated approaches. 

The obvious generalization of the Hartree-Fock approximation 
to interacting bosons would be searching for the best wavefunction 
within the subspace of permanents, 
the bosonic analogues of Slater determinants. However, as we already mentioned, 
unlike Slater determinants, permanents are  
well defined even if the single-particle wavefunctions that are 
used are not orthogonal to each other. 
Therefore the optimization procedure is not reduced to 
solving a single eigenvalue problem, which would produce a 
set of orthogonal wavefunctions, but 
becomes rather complicated, practically unfeasible, 
hence further approximations are required.\cite{anderson} 

Our simplified approach consists in adding one boson at a time with a wavefunction that is 
the ground state of a non-interacting Hamiltonian with a potential 
determined by the already added bosons. 
Specifically, we consider the not normalized $N$ boson wavefunction 
\be
\label{Psi}
|\Psi_N\rangle= \prod_{\alpha=1}^N\left(\sum_{j} \chi_{\alpha j}\,b^\dagger_j\right) |0\rangle
\ee
where $\chi_{\alpha j}, \;j=1,...,N_s$ are generically 
non-orthogonal single-particle wavefunctions. 
The first wavefunction $\chi_{1 j}$ is the  ground state 
of \eqn{Ham} with $U=0$ (no bosons are present). 
The $(M+1)$-th wavefunction is instead the ground state of 
\be
H_{app}=-\frac{t}{2} \sum_{ij} b_i^\dagger b_j + \sum_i \bigg(\epsilon_i + 
U \langle n_i\rangle\bigg)\,n_i,
\label{H-app}
\ee
where 
\be
\langle n_i \rangle=\frac{\langle \Psi_{M}|b_i^\dagger b_i|\Psi_{M}\rangle}
{\langle \Psi_{M}|\Psi_{M}\rangle},
\label{ni}
\ee
is the average density of the previously added $M$ bosons. 
We define an $M\times M$ matrix
\be
\label{Omega}
\Omega_{\alpha \beta}=\sum_{i=1}^{N_s} \chi_{\alpha i}^\dagga \chi_{\beta i}^{*}
\ee
$\alpha, \beta=1,\dots,M$, and, for each couple of lattice sites $(i,j)$, the $(M+1)\times (M+1)$ 
matrices
\begin{displaymath}
D_{ij}=\left(
\ba{ll}
\Omega& \hat\chi_{i}\\
\hat\chi_{j}^{\dagger} &\delta_{ij}
\ea
\right)
\end{displaymath}
where $\hat\chi_{i}=(\chi_{1 i},...,\chi_{M i})^t$. It follows that the mean local density 
that is required for adding the next $(M+1)$-th boson, namely Eq.(\ref{ni}), can be written as follows
\be
\label{n_exact}
\langle n_i \rangle=\frac{\textrm{Per}(D_{ii})}{\textrm{Per}(\Omega)}-1
\ee
where $\textrm{Per}(X)$ is the permanent of $X$. At each iteration, one can 
also calculate the inter-site density matrix, since 
\be
C_{ij}\equiv \frac{\langle \Psi|b_i^\dagger b_j|\Psi\rangle}{\langle \Psi|\Psi\rangle}
=\frac{\textrm{Per}(D_{ji})}{\textrm{Per}(\Omega)}-\delta_{ij},\label{C_ij}
\ee
hence investigate the eventual offset of long-range order. This procedure is iterated until the 
desired number $N$ of bosons is reached. 
We note that the method can be easily extended to study excited states -- it is sufficient 
to select at any iteration not the ground state but an excited one -- hence at finite temperature, 
even though in what follows we just focus on the lowest energy states. 

\begin{figure}[h]
\centering
\includegraphics[width=7cm]{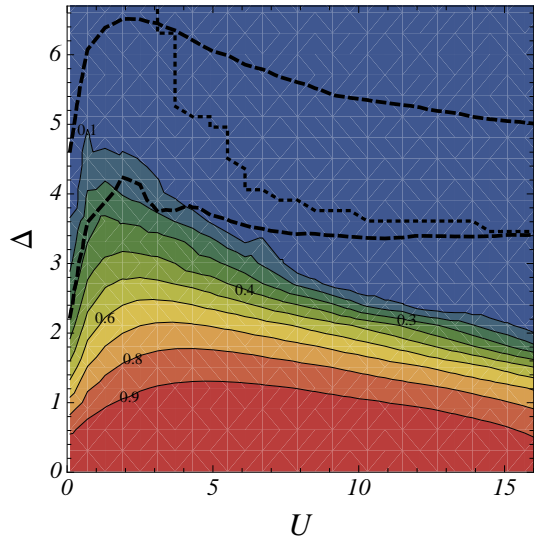}
\caption{(Color online) Contour-plot of the superfluid stiffness $\rho_{sf}$, Eq.~(\ref{rho_sf-true}), for 
a $6\times 6$ square lattice at filling $\nu=1/4$, averaging over $400$ disorder configurations. The two thick dashed lines and the dotted line show the border of the superfluid phase, by looking at some indicators related to the overlap matrix, as discussed in Sec.~\ref{comparison}.}
\label{fig1}
\end{figure}

The superfluid properties of the model can be accessed by calculating the 
superfluid stiffness $\rho_{sf}$ defined through
\be \label{rho_sf-true}
\rho_{sf}\simeq \frac{L^{2}}{N}\,  
\frac{\partial^2 E_\theta}{\partial \theta^2}\Big{|}_{\theta=0}\; ,
\ee
where $E_\theta$ is the average value of the Hamiltonian \eqn{Ham} with twisted boundary conditions 
along a given direction $\mathbf{x}$, or, alternatively, 
with hopping parameter between neighboring sites $t_{ij}= t\,e^{i\theta \vec{r}\cdot \vec{x}}$, where 
$\vec{r}=(\mathbf{R}_i-\mathbf{R}_j)/L$. In terms of permanents we calculate
\be
E_\theta=-\sum_{ij}\left(\frac{t_{ij}}{2}-\delta_{ij}(\epsilon_i-2 U)\right) C_{ij}+\frac{U}{2} \sum_i \left(\frac{\textrm{Per}(I_{iiii})}{\textrm{Per}(\Omega)}-2\right)
\ee
where $C_{ij}$ is defined in Eq.~(\ref{C_ij}) and $\textrm{Per}(I_{ijkl})= \langle \Psi|b_ib_jb_k^\dagger b_l^\dagger|\Psi\rangle$ with 
\be
I_{ijkl}=\left(
\ba{lll}
\Omega& \hat\chi_{i}& \hat\chi_{j}\\
\hat\chi_{k}^{\dagger} &\delta_{ik}&\delta_{jk}\\
\hat\chi_{l}^{\dagger} &\delta_{il}&\delta_{jl}
\ea
\right).
\ee
Upon repeating this calculation for several disorder configurations, averaging 
over them and setting equal to zero the stiffness in the region of parameters where its variance 
is greater than its average (cutting, therefore, the values of $\rho_{sf}$ statistically undetermined), 
we finally obtain the phase diagram of Fig.\ref{fig1}, which is a contour-plot of the 
averaged superfluid stiffness $\rho_{sf}$ of a $2$-dimensional model with filling fraction $\nu=N/N_s=1/4$. 
We note that in a finite system there cannot be a true gauge-symmetry breaking, hence the 
phase diagram Fig.\ref{fig1} is just an indication of what could happen in the thermodynamic limit. 
Nevertheless, the qualitative behavior that we find is physically sensible: $\rho_{sf}$ decreases 
on increasing disorder and, at fixed disorder, first increases with $U$ and then diminishes. In Fig.~\ref{fig1} and in the following figures the values of $U$ and $\Delta$ are in units of $t$.  

\begin{figure}[h]
\includegraphics[width=11cm]{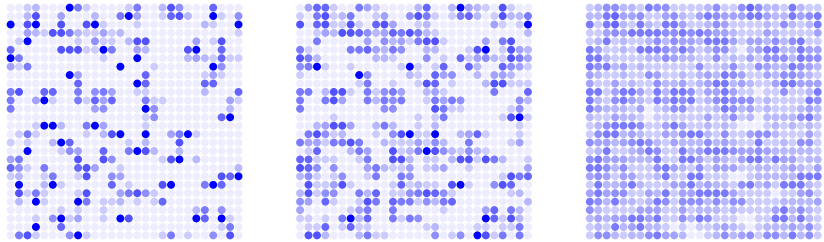}
\caption{(Color online) Contour plot of the particle density $\langle n_i\rangle$ on a $28\times 28$ lattice 
with $N=49$ bosons, i.e. filling $\nu=1/16$, at $U=10$ and for a single realization of disorder 
with $\Delta=1, 2, 3$ from right to left (i.e. crossing the transition, 
see inset of Fig.~\ref{fig.entro-stiff}). The darker the sites the higher is the density.}
\label{fig.density}
\end{figure}

We previously mentioned that a realistic view of a Bose glass is that of disconnected droplets. 
A way to confirm this idea is plotting the average density as shown in Fig.~\ref{fig.density}, 
where we used, instead of Eq.~(\ref{ni}), $\langle n_i \rangle\simeq \sum_{\alpha}^N |\chi_{\alpha i}|^2$, 
which is a good approximation for local densities, in order to computationally reach larger system sizes.
We find that, at large disorder, bosons are indeed concentrated into droplets whose magnitude increases 
on decreasing disorder until a percolating cluster appears, which must presumably signal the onset 
of superfluid.  

\subsection{Comparison with the overlap matrix method}
\label{comparison}
Let us compare the transition obtained by the superfluid stiffness with that given by the overlap matrix criterium.
In our case, given the many-body wavefunction Eq.~\eqn{Psi}, 
the overlap matrix $\Omega$ is that 
defined in Eq.~(\ref{Omega}) with $\psi_{\alpha i}=\chi_{\alpha i}$. 
The action is given by Eq.~(\ref{act}) and the saddle point equation reads 
as in Eq.~\eqn{sp}, which has non trivial solution if $\Omega$ 
has eigenvalue 2. 

In Fig.~{\ref{fig1}} we plot a dotted line below which $(\Omega-2\mathbb{1})$, 
or equivalently, $({\sf O}-2\mathbb{1})$, 
has both positive and negative eigenvalues for any disorder configuration,
implying that the saddle point equation has always non trivial solutions. 
Above that line, instead, for some disorder configurations all the eigenvalues 
of $\Omega$ are smaller than $2$. 

On the same figure, Fig.~{\ref{fig1}}, we plot two thick dashed lines which 
correspond to $\overline{\cal F}=2$ (the upper line), namely, when the value 
of ${\cal F}$ as defined in Eq.~(\ref{cal_F}), averaged over $400$ disorder 
configurations, equals $2$, and ${\it min}({\cal F})=2$ (lower line), 
namely, when the minimum value of ${\cal F}$ among the disorder 
configurations equals $2$. 
Below those lines the corresponding quantities exceed the threshold value.
Notice that a non trivial solution of the saddle point equation occurs also 
for large disorder and weak interaction (the dotted line keeps on growing 
decreasing $U$) while ${\cal F}$, which detects the long range order (dashed lines), follows correctly the stiffness behavior also for small interaction.

In summary, we find that the superfluid properties of the many-body 
wavefunction \eqn{Psi} can be
related to the overlap matrix $\Omega$ between the single-particle 
wavefunctions $\chi_{\alpha i}$.
This result also clarifies why,
even though each wavefunction $\chi_{\alpha i}$ is the lowest energy 
solution of the Schr{\"o}dinger
equation of a particle in a disordered potential, hence would be always 
localized in two dimensions,
and also in higher dimensions if, as presumably is the case, it lies in the 
Lifshitz tail,
nevertheless the permanent built with them could still be superfluid.

\begin{figure}[h]
\includegraphics[width=8.2cm]{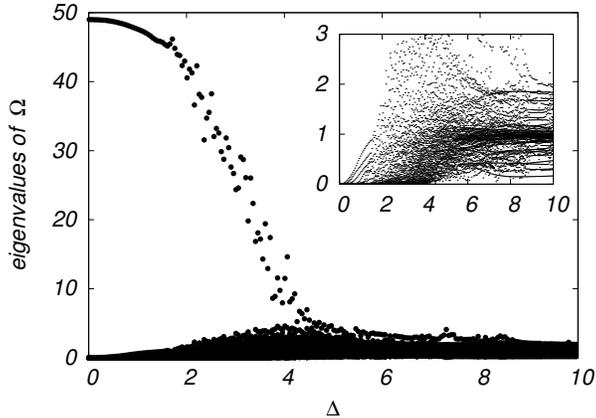}
\caption{Eigenvalues of $\Omega$ as functions of $\Delta$, at a given configuration of disorder, for $49$ bosons on a square lattice with $256$ sites, namely always for filling $1/4$, at $U=10$. The inset magnifies the lower part of the main plot.}
\label{levels}
\end{figure}

A final comment. We optimized the wavefunction by explicitly 
evaluating permaments. This is in general very cumbersome, not much simpler 
than an exact numerical solution of the problem, which is the reason why 
our simulation size is small. However, we could adopt an oversimplified 
approach and evaluate the Hartree potential in Eq.~\eqn{H-app} for the 
$(M+1)$-th boson using 
\be
\label{apxn}
\langle n_i\rangle \simeq \sum_{\alpha =1}^M\,|\chi_{\alpha i}|^2,
\ee
as if the already present $M$ bosons were distinguishable.  
This approximation simplifies a lot the procedure to determine the 
single-particle wavefunctions, 
which can be pushed to very large system sizes. These single-particle 
wavefunctions can then be used to construct the permanent, whose superfluid 
properties can be assesed by the overlap matrix criterium. Even though such a 
procedure is hardly justifiable from the variational point of view, it 
provides a phase diagram qualitatively correct. Using this approximation one 
can study the spectrum of the overlap matrix $\Omega$ for larger system, 
finding that, indeed, the transition is characterized by the fact that the
greatest eigenvalue of $\Omega$ starts growing towards value $N$, the number 
of bosons, for $\Delta$ going to zero. From Fig.~\ref{levels} one can see 
clearly, in fact, that a single eigenvalue of $\Omega$ (although quite noisy, 
it is a single level) separates from the others at $\Delta \approx 4$, 
approximately at the same value of disorder as that obtained for a smaller 
system size with $\langle n_i\rangle$ given by (\ref{n_exact}) (cfr. the transition line at $U=10$ in Fig.~\ref{fig1}). 
The other eigenvalues (see the inset of Fig.~\ref{levels}) accumulate around zero below the transition, and 
around one above it, at least for filling not greater than one, as in our case.
  
\section{Spatial entanglement entropy}
\label{sec.entropy}
Another quantity which may be interesting to look at is the single site entanglement entropy.
We define as $\rho_n(i)$ the probability to have $n$ bosons at site $i$, which must trivially satisfy 
$\sum_{n=0}^N\,\rho_n(i)=1$, where $N$ is the total number of bosons. The single-site entropy $S_i$  
is thus defined through
\be
S_i  = - \sum_{n=0}^N\,\rho_n(i)\,\ln\rho_n(i).\label{def:S_i}
\ee
In a disordered system it is also convenient to define its probability distribution through 
\be
P(S) = \langle\, \frac{1}{N_s}\sum_i\,\delta\left(S-S_i\right)\,\rangle_{disorder},\label{def:P(S)}
\ee
which is obtained considering all sites and all disorder configurations. 
In spite of its simple definition, we are going to show that the single-site entanglement entropy, 
and especially its distribution probability, is a quite enlighting quantity 
in the presence of both disorder and interaction. 

\subsection{A limiting case}
We start considering the case where all bosons condense into a single state, i.e. 
$\chi_{\alpha\, j}=\zeta_j$, $\forall \alpha$, with the normalization 
condition $\sum_{j=1}^{N_s}\zeta_j=1$. In this case $\Per(\Omega)=N!$ and the state Eq.~(\ref{Psi}), 
including the normalization factor $1/\sqrt{\Per(\Omega)}$ can be decomposed in the site Fock basis as
\be
|\Psi\rangle=\frac{1}{\sqrt{N!}}\left(\sum_j \zeta_j b_j^\dagger\right)^N|0\rangle=
\sum_{\{n\}}\frac{\sqrt{N!}}{\prod_i n_i!}\prod_{j=1}^{N_s}\zeta_j^{n_j}(b_j^\dagger)^{n_j}|0\rangle.
\ee 
The sum runs over $\{n\}=(n_1,n_2,...,n_{N_s})$, the configurations of occupation numbers 
with the constraint $\sum_i n_i=N$. 
We calculate the reduced density matrix $\hat\rho^\ell$ by partitioning the sites in two blocks: 
$[1,\ell]$ and $[\ell,N_s]$, and tracing out the sites belonging to the second block:
\be
\hat\rho^\ell=\sum_{n_{\ell+1},..,n_{N_s}}\frac{1}{\prod_{i=\ell+1}^{N_s}n_i!} 
\langle 0| \prod_{i=\ell+1}^{N_s}(b_i)^{n_i}|\Psi\rangle\langle\Psi| \prod_{i=\ell+1}^{N_s}(b_i^\dagger)^{n_i}|0\rangle.
\ee
This is a diagonal $\frac{(N+1)!}{\ell!(N+1-\ell)!}\times\frac{(N+1)!}{\ell!(N+1-\ell)!}$ matrix. 
For each diagonal element $\rho_{n_1,...,n_\ell}$, corresponding to the configuration of the occupation 
numbers $(n_1,n_2,...,n_\ell)$ of the $\ell$ sites that are not traced out,  we obtain the 
following expression
\be
\rho_{n_1,...,n_\ell}=\frac{N!}{\left(N-\sum_{i=1}^\ell n_i\right)!\prod_{i=1}^\ell n_i!}
\left(1-\sum_{i=1}^\ell |\zeta_i|^2\right)^{\left(N-\sum_{i=1}^\ell n_i\right)} \prod_{i=1}^\ell |\zeta_i|^{2n_i}. 
\ee
From now on we shall focus on the simplest partitioning, keeping only one site and 
tracing over all the others, i.e. $\ell=1$.
In this case we get a very simple binomial expression of the reduced density matrix 
$\rho_n\equiv \rho_{n}(1)$
\be
\label{rho}
\rho_n=\left(
\ba{c}
N\\n
\ea
\right)
\left(1-|\zeta_1|^2\right)^{\left(N-n\right)}|\zeta_1|^{2n}. 
\ee
It is straightforward to check that $\sum_n \rho_{n}=1$. The reduced density matrix is fully local; 
it depends only on the value of the wavefunction $\zeta$ on that site. 
We can now calculate the entanglement von Neumann entropy at site $1$, $S_1$, through \eqn{def:S_i}.  
From Eqs.~(\ref{rho}) and (\ref{def:S_i}) we find that for any fixed value of $|\zeta_1|\in (0, 1)$ and for large $N$, the asymptotic  value of $S_1$ is
\be
\label{Smax}
S_{1}\xrightarrow[N\gg 1]{} S_{loc}\equiv \frac{1}{2}\ln N +A, 
\ee
with $A=\frac{1}{2}\left\{1+\ln\big[2\pi|\zeta_1|^2(1-|\zeta_1|^2)\big]
\right\}$. At the two extremes, $|\zeta_1|=0, 1$, we have $S_1=0$.  

Let us consider first the non-interacting case with disorder. All bosons condense into a localized 
wavefunction. Within the localization region $S_i\sim \ln N$, while outside $S_i=0$. 
We conclude that the probability distribution \eqn{def:P(S)} 
becomes peaked at the single value $S=0$ in the thermodynamic limit, 
$N_s\to\infty$, where the localization region has zero measure with respect to 
the whole system.

\begin{figure}[h]
\includegraphics[width=8cm]{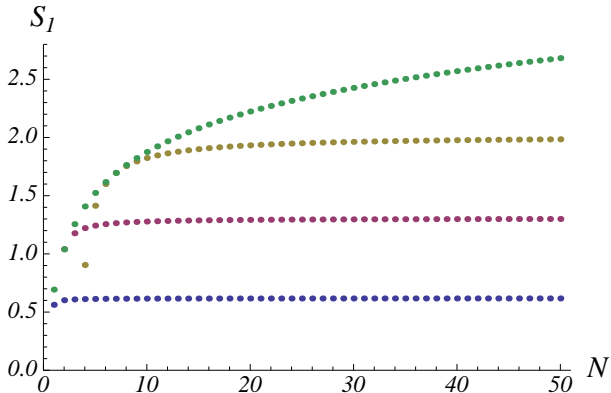}
\caption{(Color online) Single site entanglement entropies obtained by Eqs.~(\ref{rho}), (\ref{def:S_i}). 
Upper dot-line: the maximum entropy, reached when $|\zeta_1|^2=1/2$. 
Upper middle dot-line: entropy for the superfluid with $\nu=7/2$ case, i.e. for $|\zeta_1|^2=7/(2N)$.
Lower middle dot-line: entropy for the superfluid in the commensurate $\nu=1$ case, i.e. 
for $|\zeta_1|^2=1/N$. Lower dot-line: the superfluid entropy for $|\zeta_1|^2=1/(4N)$, i.e. $\nu=1/4$.}
\label{fig.entropia}
\end{figure}

On the contrary, without disorder and deep in the superfluid phase, small $U$, we expect that 
\be
|\zeta_1|^2=\frac{1}{N_s}=\frac{\nu}{N}
\ee
where the filling fraction $\nu=N/N_s$. In this case, the single-site entropy  
in the thermodynamic limit is finite, site-independent and depends exclusively on the filling fraction $\nu$:
\be
\label{Ssf}
S_{1}\xrightarrow[N\rightarrow \infty]{} S_{SF}\equiv \nu\left(1-\ln \nu \right)+
e^{-\nu}\sum_{n=0}^{\infty}\frac{\nu^n}{n!}\,\ln n! 
\ee
The sum Eq.~(\ref{Ssf}) converges for any value of $\nu$. As one can see 
from Fig.~\ref{fig.entropia}, 
the thermodynamic limit Eq.~(\ref{Ssf}) is approached already with few bosons, 
for small $\nu$. 
In the same figure we also draw the maximum entropy line, Eq.~(\ref{Smax}), which indeed lies above all curves. 

In conclusion, we expect that the probability distribution 
of the single-site entropy is peaked at a finite value deep in the superfluid phase while 
at value zero deep in the Bose-glass phase. The obvious question is what happens in between. 

\subsection{Single-site entropy across the superfluid-to-Bose glass transition}
\label{with_disorder}

\begin{figure}[h]
\includegraphics[width=7.5cm]{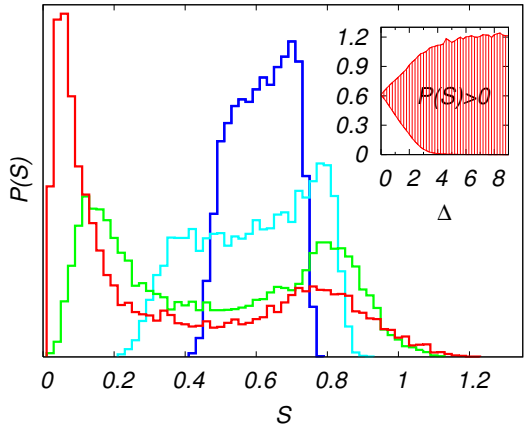}
\caption{(Color online) Probability distribution of entropy for $U=10$ and for different values of disorder strength 
($\Delta =1, 2, 4, 6$). The plot has been made after $10^4$ realizations of disorder in a system 
of $N=4$ bosons on a $4\times 4$ square lattice. In the inset: the fluctuations of $S$ increase 
with the strength of disorder and saturate to the maximum value of entropy for $4$ bosons which is $\simeq 1.4$}
\label{fig.istogram}
\end{figure}

\begin{figure}[h]
\includegraphics[width=8.cm]{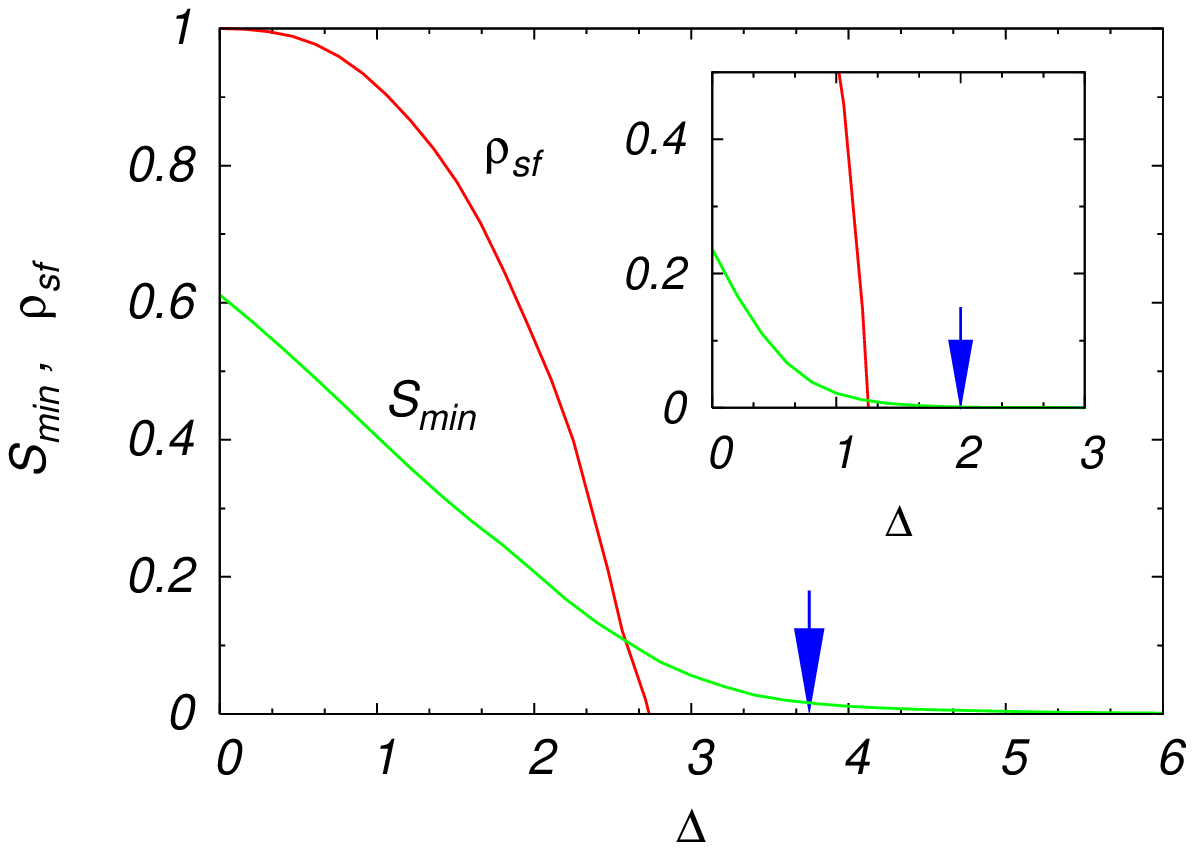}
\caption{(Color online) The minimum site entanglement entropy $S_{min}$ (light-green lines) compered 
with the superfluid stiffness $\rho_{sf}$ (dark-red lines) as functions of the disorder, 
for $U=10$ and for filling fraction $\nu =1/4$ (main plot) and $\nu=1/16$ (inset). 
The stiffness, $\rho_{sf}$, is taken as an average among $400$ configurations of disorder 
for a system of $N=9$ bosons on a square lattice of $6\times 6$ ($\nu=1/4$, main plot) 
and $12\times 12$ ($\nu=1/16$, inset). The minimum entropy $S_{min}$ is obtained taking the 
minimum values of $S$ among $10^4$ configurations of disorder, and for $N=4$ on a $4\times 4$ 
($\nu=1/4$, main plot) and $8\times 8$ ($\nu=1/16$, inset) lattices. 
The blue arrows point at the transitions, obtained from the overlap matrix method as 
explained in Sec.~\ref{sec.path}.}
\label{fig.entro-stiff}
\end{figure}

Let us now consider the state Eq. (\ref{Psi}), including the normalization factor $1/\sqrt{\Per(\Omega)}$, 
which in the Fock basis of sites reads
\be
\label{Fock}
|\Psi\rangle=\frac{1}{\sqrt{\Per(\Omega)}}\sum_{\{n\}}\sum_{\{\{p_j\}\}}\prod_{j=1}^{N_s} \prod_{\alpha=1}^{n_j}
\chi_{p_j(\alpha)\, j} (b_j^\dagger)^{n_j}|0\rangle. 
\ee
The sum runs over $\{n\}=(n_1,n_2,...,n_{N_s})$, the configurations of occupation numbers 
with the constraint $\sum_i n_i=N$, and over $\{p_j\}$ with 
$\sum_{\{\{p_j\}\}}=\sum_{\{p_1\}}\sum_{\{p_2\}\neq \{p_1\}}\sum_{\{p_3\}\neq {\{p_1\},\{p_2\}}}...$ where 
$p_j(\alpha)$ is the particle label which takes $n_j$ integer values among $N$ values.
The reduced density matrix for a single site is found to be 
\be
\rho_n=\frac{n!}{\Per(\Omega)}\sum_{n_2,..,n_{N_s}}\prod^{N_s}_{i=2} n_i!
\Bigg|\sum_{\{\{p_j\}\}}\prod_{j=1}^{N_s} \prod_{\alpha=1}^{n_j}\chi_{p_j(\alpha)\, j}{\Bigg{|}}^2
\ee
The explicit expressions of the reduced density matrix for $N=2$ and $N=4$ 
are given in Appendix~\ref{example}. In what follows we will consider the simple case of $N=4$ bosons. 
In Fig.~\ref{fig.istogram} we show the probability 
distribution of the single site entanglement entropy , which, in the absence of disorder, 
is peaked around $S\simeq 0.62$, broadens under the action of disorder and finally, when 
disorder becomes strong, develops again a peak but at $S=0$. 
We observe that the probability of having zero entropy becomes non-zero only above a disorder threshold 
that occurs approximately when also the stiffness vanishes, as one can see from Fig.~\ref{fig.entro-stiff}. 

We have found that in our finite size simulation the crossover between the two limiting 
behaviors, $P(S)$ peaked either at $S\not =0$, deep in the superfluid, or close to $S=0$, deep in the 
Bose glass, is continuous. We have also noticed that the point at which $P(S)$ becomes finite at $S\simeq 0$ 
seems to coincide with the point where the stiffness becomes very small, which could be 
taken as the finite size signal of the superfluid-to-Bose glass transition.  
This suggests that the behavior of $P(S=0)$ could be used as well to identify the 
superfluid-to-Bose glass transition in alternative to the superfluid stiffness 
in any finite-size numerical simulations. 
Moreover, we note, looking at Fig.~\ref{fig.istogram}, that, close to the transition ($\Delta\approx 4$), the 
profile of the distribution (the green curve) is almost an equally weighted bimodal, peaked either 
at small $S$ (insulating regions) 
and at 
$$S\approx S_{SF}(\nu_{eff})$$ (superfluid regions)
given by Eq. (\ref{Ssf}) 
where $\nu$ is replaced by $\nu_{eff}= \nu/p_c$. 
The numerical result for the second peak close to the transition 
($S\approx 0.8$, for $\nu=1/4$), is consistent 
with $p_c\approx 0.59$ which is 
the site percolation threshold for a square lattice. We have checked this 
result also with other filling fractions.
This finding is in a nice agreement with the percolating droplets scenario.

\section{Conclusions}
\label{sec.conclusions}
We have considered a two-dimensional Bose-Hubbard Hamiltonian in the presence of disorder, 
and constructed a trial many-body wavefunction by solving a single-particle problem 
for each boson at a time that feels the effect of all the others as an effective potential.
By this wavefunction we have studied the transition between the superfluid and the Bose glass. 
Beside the vanishing stiffness, we find that the transition can be characterized by the eigenvalues  
of the matrix of the overlaps between the single-particle wavefunctions. 
Another quantity we propose to look at is the single-site entanglement entropy. 
In particular, we argue that when the probability to measure zero entropy 
becomes finite, $P(S=0)>0$, 
the superfluid start to vanish and the Bose glass phase sets in. 
The bimodal entropy distribution is, then, the signature of lack of 
percolation among superfluid clusters.

\appendix
\section{Derivation of the action}
\label{action}
Let us consider the following wavefunction
\be
\label{wavefn}
|\Psi\rangle=\prod_\alpha \left(\sum_i\psi_{\alpha i}\,b_i^\dagger\right)^{n_\alpha}|0\rangle,
\ee
which generalizes Eq.~(\ref{wavef}) where the integers $n_\alpha$ are all 
equal to $1$. Let us define also the overlap matrix $\Omega_{\alpha\beta}=
\sum_{i}\psi_{\alpha i}\psi^{*}_{\beta i}$ which is Hermitian and positive defined, such that can be parametrized as $\Omega=\psi\psi^\dagger=V^\dagger |\lambda|^2V$, in terms of a unitary matrix $V$. Let us suppose now that another unitary matrix $U$ exists so that we can write $\psi=V^\dagger \lambda U$, and define the following bosonic operators 
\bea
\label{ba}
&&b^\dagger_a=\sum_i U_{a i}b^\dagger_i\\ 
&&b^\dagger_\alpha=\sum_a V_{\alpha a}^\dagger b^\dagger_a. 
\eea
Defining also 
\be
e^{\cal T}\equiv e^{\sum_a b^\dagger_a b_a\ln\lambda_a }
\ee
one can then verify that
\be
\sum_i \psi_{\alpha i}b^\dagger_i=\sum_{a}V^\dagger_{\alpha a}\lambda_a b^\dagger_a=e^{\cal T} b^\dagger_\alpha e^{-\cal T}
\ee
where we have used the Hausdorff relation 
\be
\lambda_a b^\dagger_a=e^{\cal T} b^\dagger_a e^{-\cal T}. 
\label{hausdorff}
\ee
Therefore Eq.~(\ref{wavefn}) can be written as follows
\be
|\Psi\rangle= e^{\cal T} \prod_\alpha \left(b^\dagger_\alpha\right)^{n_\alpha}|0\rangle
\ee
Moreover, one can check that
\be
\label{eTeT}
e^{\cal T^\dagger}e^{\cal T}=e^{\sum_a b^\dagger_a b_a\ln|\lambda_a|^2}=\frac{1}{\pi}\int \prod_a dz_a dz^*_a e^{-\sum_a|z_a|^2}e^{\sum_a z_a v^*_a b_a^\dagger}e^{\sum_a z^*_a v_a b_a}
\ee
with $|\lambda_a|^2=1+|v_a|^2$. Now, in order to rewrite Eq.~(\ref{eTeT}) on the basis of $b_\alpha$ instead of $b_a$, we have to define 
${\hat v}_{\alpha i}=\sum_a V^\dagger_{\alpha a} v_a U_{a i}$ and $\xi_\alpha=\sum_{i\,a} {\hat v}_{\alpha i}U^\dagger_{i a} z^*_a$, and noticing that ${\hat v}{\hat v}^\dagger=\Omega -\mathbb{1}$, we finally have
\be
\langle \Psi|\Psi\rangle=\int \prod_{\alpha} \frac{d \xi_{\alpha} d \xi_{\alpha}^*}{\pi}e^{-\xi^\dagger(\Omega -\mathbb{1})^{-1}\xi}\,
\langle 0|\prod_{\beta} (b_{\beta})^{n_{\beta}}\, e^{\sum_\alpha\xi_{\alpha}^* b^\dagger_{\alpha}}e^{\sum_\alpha\xi_{\alpha} b_{\alpha}} \prod_{\gamma} (b_{\gamma}^\dagger)^{n_{\gamma}}|0\rangle            
\ee
Expanding the exponents and using the commutation relations, in particular the equality $b^m b^{\dagger n}|0\rangle =\frac{n!}{(n-m)!}b^{\dagger (n-m)}|0\rangle$ for $m\le n$, one can verify that
\be
\langle 0|\prod_{\beta} (b_{\beta})^{n_{\beta}}\, e^{\sum_\alpha\xi_{\alpha}^* b^\dagger_{\alpha}}e^{\sum_\alpha\xi_{\alpha} b_{\alpha}} \prod_{\gamma} (b_{\gamma}^\dagger)^{n_{\gamma}}|0\rangle = \prod_\alpha 
\sum^{n_\alpha}_{m=0} \frac{(n_\alpha!)^2}{(m!)^2(n_\alpha-m)!}|\xi_\alpha|^{2m}
\ee
equal to $\prod_\alpha n_\alpha!\, L^0_{n_\alpha}(-|\xi_\alpha|^2)$, using the definition of the generalized Laguerre polynomials, $L^k_{n}(z)=\sum_{m=0}^n\frac{(n+k)!}{m!(k+m)!(n-m)!}(-z)^m$, which can be written also in terms of confluent hypergeometric functions, $L^k_{n}(z)=\frac{(k+n)!}{k!n!}\,\phantom{}_1F_1(-n, k+1 ; z)$. As a final result the norm of (\ref{wavefn}) can be written as an integral, Eq.~(\ref{norm}), with the action
\be
\label{actn}
{\cal{S}}(\xi,\xi^\dagger)=\xi^\dagger(\Omega-{\mathbb{1}})^{-1}\xi-\sum_\alpha
\,\ln\bigg[n_\alpha!\,L^0_{n_\alpha}\left(-|\xi_\alpha|^2\right)\bigg].
\ee
If $n_\alpha=1$, $\forall \alpha$, since $L^0_{1}(-|\xi_\alpha|^2)=(1+|\xi_\alpha|^2)$, the action 
reduces to Eq.~(\ref{act}).
The saddle point equation reads then
\be
\xi_\alpha = \sum_{\beta, n_\beta>0}\, \left(\Omega_{\alpha\beta}-
\delta_{\alpha\beta}\right)\,
\frac{\displaystyle L^1_{(n_\beta-1)}\left(-|\xi_\beta|^2\right)}
{\displaystyle L^0_{n_\beta}\left(-|\xi_\beta|^2\right)}
\,\xi_\beta, \label{sp-eq2}
\ee
which reduces to Eq.~(\ref{sp-eq}), for all $n_\alpha=1$. Since, for $|\xi_\beta|^2\ll 1$, we can expand $L^k_n\left(-|\xi_\beta|^2\right)\simeq
\frac{(n+k)!}{n!\, k!}\bigg(1+\frac{n}{k+1}|\xi_\beta|^2\bigg)$, 
the eigenvalue equation, when discarding $O(\xi_\beta^3)$ terms in the 
r.h.s. of Eq.~(\ref{sp-eq2}), becomes 
\be
\xi_\alpha
=\sum_{\beta}\, \left(\Omega_{\alpha\beta}-\delta_{\alpha\beta}\right)
n_\beta\,\xi_\beta,
\label{sp2}
\ee
which generalizes Eq.~(\ref{sp}) for arbitrary $n_\beta$.
\section{Reduced density matrix: two simple cases}
\label{example}
Here we are going to show within a toy model made of $2$ bosons arranged on $3$ sites how disorder and interaction can conspire to enhance the bipartite entanglement entropy. With two bosons $N=2$ and without disorder, in the delocalized phase, $|\zeta_1|^2=1/3$, from Eqs. (\ref{rho}), (\ref{def:S_i}), we get the following single site entropy
\be
\label{S2bosons}
S= 2\ln 3 -\frac{16}{9}\ln 2  \simeq 0.96
\ee
What we shall be seeing in the following is that this value can be overcome introducing a suitable amount of disorder.\\
If we now take Eq.(\ref{Fock}) as the many body wave function, we get, for $N=2$ bosons and generic $N_s$ sites, the following three diagonal elements of the reduced density matrix
\bea
\label{r0}
\rho_0&=&\frac{1}{\Per(\Omega)}\sum_{i=2}^{N_s}\left(2|\chi_{1 i}\chi_{2 i}|^2+\sum_{j> i}^{N_s}|\chi_{1 i}\chi_{2 j}+\chi_{2 i}\chi_{1 j}|^2\right)\\
\label{r1}
\rho_1&=&\frac{1}{\Per(\Omega)}\sum_{i=2}^{N_s}|\chi_{1 1}\chi_{2 i}+\chi_{2 1}\chi_{1 i}|^2\\
\label{r2}
\rho_2&=&\frac{2}{\Per(\Omega)}|\chi_{1 1}\chi_{2 1}|^2
\eea
Let us now consider $N_s=3$ and a simple on site disorder such that $\epsilon_i=\pm \Delta$. We can have therefore $8$ possible configurations of disorder. 
For $U\gg\Delta$, by semiclassical considerations and using Eqs.~(\ref{r0}-\ref{r2}), we can have probability $\sim 1/4$ to have $S\simeq 0$, i.e.  $P(S\simeq 0)\simeq 1/4$ while $P(S\simeq \ln 2 )\simeq 1/2$ and $P(S\simeq 0.96)\simeq 1/4$, the same entropy as without disorder. 
For $U\ll\Delta$, instead, we have $P(S\simeq 0)\simeq 1/2$, $P(S\simeq 0.96)\simeq 1/4$ and $P(S\simeq \frac{3}{2}\ln 2 \simeq 1.04)\simeq 1/4$, namely we can have a sizable probability to have maximum entropy for two bosons, so to exceed the value in Eq. (\ref{S2bosons}). The disorder, therefore, peaks the probability distribution at zero while widening the entropy fluctuations.\\
In general for filling $\nu< 1$, the superfluid single site entropy is $S\simeq \nu(1-\ln \nu )$, as shown in the text. By introducing strong enough disorder there is the possibility for a fraction $k$ of the total sites to have very large local energies which make those sites inaccessible for the particles. This induces a larger effective filling, $\frac{\nu}{1-k}$, and consequently enhances the entropy.
 
For the sake of completeness we hereafter report the form of the reduced density matrix for $N=4$ bosons, used in the paper in Sec.~\ref{with_disorder}. To simplify notation we define  
the matrix
\be
{\cal L}^{ijkl}=\left(
\ba{cccc}
\chi_{1 i}&\chi_{1 j}&\chi_{1 k}&\chi_{1 l}\\
\chi_{2 i}&\chi_{2 j}&\chi_{2 k}&\chi_{2 l}\\
\chi_{3 i}&\chi_{3 j}&\chi_{3 k}&\chi_{3 l}\\
\chi_{4 i}&\chi_{4 j}&\chi_{4 k}&\chi_{4 l}
\ea
\right)
\ee
so that the single site reduced density matrix can be written in the following form:\\
\bea
\rho_0&=&\frac{1}{\Per(\Omega)}\sum_{i=2}^{N_s}\left\{\sum_{j>i}^{N_s}\sum_{k>j}^{N_s}\sum_{l>k}^{N_s}|\Per({\cal L}^{ijkl})|^2 +\frac{1}{2!}\sum_{j\neq i}^{N_s}\sum_{k\neq i,j}^{N_s}{|\Per({\cal L}^{iijk})|^2}\right.\\
\nonumber &&\left.+\frac{1}{2!2!}\sum_{j>i}^{N_s}{|\Per({\cal L}^{iijj})|^2}+\frac{1}{3!}\sum_{j\neq i}^{N_s}{|\Per({\cal L}^{iiij})|^2}+\frac{1}{4!}{|\Per({\cal L}^{iiii})|^2}\right\}\\
\rho_1&=&\frac{1}{\Per(\Omega)}\sum_{i=2}^{N_s}\left\{\sum_{j>i}^{N_s}\sum_{k>j}^{N_s}|\Per({\cal L}^{1ijk})|^2
+\frac{1}{2!}\sum_{j\neq i}^{N_s} {|\Per({\cal L}^{1iij})|^2}+\frac{1}{3!}{|\Per({\cal L}^{1iii})|^2}\right\}\\
\rho_2&=&\frac{1}{\Per(\Omega)}\sum_{i=2}^{N_s}\left\{\frac{1}{2!}\sum_{j>i}^{N_s}{|\Per({\cal L}^{11ij})|^2}+\frac{1}{2!2!}{|\Per({\cal L}^{11ii})|^2}\right\}\\
\rho_3&=&\frac{1}{\Per(\Omega)}\,\frac{1}{3!}\sum_{i=2}^{N_s} {|\Per({\cal L}^{111i})|^2}\\
\rho_4&=&\frac{1}{\Per(\Omega)}\,\frac{1}{4!}{|\Per({\cal L}^{1111})|^2}
\eea
\\
From equations above one can easily guess the form of the reduced density matrix for a generic value of $N$.
\bibliographystyle{apsrev}

\end{document}